\documentclass[a4paper]{svmult}
\usepackage{times}
\usepackage[dvips]{graphicx,epsfig}
\usepackage{amsmath,amsfonts,amssymb}
\usepackage{cite,url}

\begin{document}

\frontmatter

\mainmatter

\title*{Deeply Virtual Compton Scattering off Nuclei}

\author{\underline{Eric Voutier}}

\titlerunning{Deeply Virtual Compton Scattering off Nuclei}
\authorrunning{Eric Voutier}

\toctitle{Deeply Virtual Compton Scattering off Nuclei}
\tocauthor{Eric Voutier}

\institute{Laboratoire de Physique Subatomique et de Cosmologie\\
IN2P3/CNRS - Universit\'e Joseph Fourier - INP \\
53 rue des Martyrs\\
38026 Grenoble cedex, France}{}

\maketitle

\begin{abstract}
Deeply virtual Compton scattering (DVCS) is the golden exclusive channel for 
the study of the partonic structure of hadrons, within the universal framework 
of generalized parton distributions (GPDs). This paper presents the aim and 
general ideas of the DVCS experimental program off nuclei at the Jefferson 
Laboratory. The benefits of the study of the coherent and the incoherent 
channels to the understanding of the EMC (European Muon Collaboration) 
effect are discussed,
along with the case of nuclear targets to access neutron GPDs.
 
\end{abstract}

             
\hyphenation{cor-res-pon-ding}

\section{Introduction}

The perspective of an access to the partonic structure of hadrons in the 
scaling regime of exclusive processes, particularly the contribution of the 
quark orbital momentum to the nucleon spin~\cite{Ji97}, sustains an important 
theoretical and experimental activity. Pioneer measurements at 
HERMES~\cite{Air01} and CLAS~\cite{Ste01} have demonstrated the relevance of the 
deeply virtual Compton scattering (DVCS) process to these studies. Together 
with recent dedicated DVCS experiments~\cite{{Mun06},{Gir08}} at the Jefferson 
Laboratory (JLab) supporting the scaling of the cross section at transferred 
momentum as low as 2~GeV$^2$, the road towards a systematic investigation of 
novel features of the inner structure of hadrons is opened. 

Generalized Parton Distributions (GPDs)~\cite{{Mul94},{Rad97}} provide a 
powerful and appealing framework for the description of the partonic structure 
of hadrons. GPDs represent the interference between amplitudes corresponding 
to different quantum states and describe the correlations between quarks, 
anti-quarks, and gluons. They can be interpreted as the transverse 
distribution of partons carrying a certain longitudinal momentum fraction 
of the hadron~\cite{{Bur00},{Ral02},{Die02},{Bel02}}, providing then a natural 
link with the transverse degrees of freedom. Therefore, GPDs unify in the 
same framework electromagnetic form factors, parton distributions, and the 
spin of the nucleon~\cite{Die03}.

The observation by the European Muon Collaboration (EMC) of a deviation of 
the deep inelastic structure function of a nucleus from the sum of the 
structure functions of the free nucleons~\cite{Aub83} (EMC effect), shows that 
the nuclear environment has a significant impact on the partonic structure of 
nucleons. Many years of theoretical and experimental efforts have led to the 
identification of four different kinematical regimes, however without 
providing a clear physical interpretation of this phenomenon. GPDs bring a 
new and specific light on the understanding of the EMC effect by allowing the  
investigation of the role of parton correlations as well as the parton 
distributions of bound nucleons, via the deeply virtual Compton scattering
(DVCS) process.
    
\section{Principle of Experiments}
 
\begin{figure}[h]\centering
\includegraphics[width=59mm]{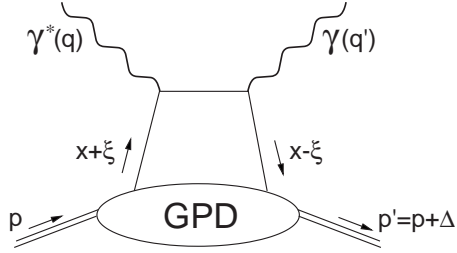}
\caption{Lowest order (QCD) amplitude for the virtual Compton process. The 
momentum four-vectors of the incident and scattered photon are $q$ and $q'$, 
respectively. The momentum four-vectors of the initial and final proton are 
$p$ and $p'$, with $\Delta$=$(p'-p)$=$(q-q')$. The DIS (deep inelastic
scattering) scaling variable is
$x_{\rm B}$=$Q^2/(2 p \cdot q)$ and the DVCS scaling variable is $\xi
$=$x_{\rm B}/(2-x_{\rm B})$. In light cone coordinates defined by 
$P$=$(p+p')/2$, the initial and final momentum of the photons are $-2\xi$ 
and $0$, respectively.}
\label{fig:hdbg}
\end{figure}

GPDs are universal non-perturbative objects entering the description of hard
scattering processes. They are defined for each quark flavor $f$ and gluon, 
and correspond to the amplitude for removing a parton of longitudinal 
momentum fraction $x+\xi$ and restoring it with momentum fraction $x-\xi$ 
(fig.~\ref{fig:hdbg}). In this process, the nucleus receives a momentum 
transfer $t=\Delta^2$ which transverse component $\Delta_{\perp}$ is Fourier 
conjugate to the transverse position of partons.

DVCS, corresponding to the absorption of a virtual photon by a quark  
followed quasi-instantaneously by the emission of a real photon, is the 
simplest reaction to access GPDs. In the Bjorken regim, $-t \ll Q^2$ and $Q^2$ 
much larger than the quark confinement scale, the leading contribution to the 
reaction amplitude is represented by the so-called handbag diagram 
(fig.~\ref{fig:hdbg}) which figures the convolution of a known $\gamma^{\ast} 
q \to \gamma q$ hard scattering kernel with an unknown soft matrix element 
describing the partonic structure of the nucleus parametrized by 
GPDs~\cite{{Ji98},{Col99}}. Consequently, GPDs ($H^f$) enter the reaction 
cross section through the Compton form factor $\cal H$ which involves an 
integral over the intermediate quark propagators
\begin{eqnarray}
{\cal H} & = & \sum_f e_f^2 \, \, {\cal P} \int_{-1}^{+1} dx \, \left( \frac{1}{x-\xi} + 
\frac{1}{x+\xi} \right) H^f(Q^2,x,\xi,t) \label{eq:cff} \\
& \phantom{-} & \hspace*{70pt} -i \pi \sum_f e_f^2 \left[ H^f(Q^2,\xi,\xi,t) - 
H^f(Q^2,-\xi,\xi,t) \right] \nonumber 
\end{eqnarray}
$e_f$ being the electric charge of the considered quark flavor in unit of the 
elementary charge.
 
In addition to the DVCS amplitude, the cross section for electroproduction of 
photons gets contributions from the Bethe-Heitler (BH) process where the real 
photon is emitted by the initial or final lepton, leading to
\begin{equation}
\frac{d^5 \sigma}{dQ^2 dx_B dt d\phi_e d \varphi} = {\cal T}_{BH}^2 + {\vert 
{\cal T}_{DVCS} \vert}^2 + 2 \, {\cal T}_{BH} \Re e\{ {\cal T}_{DVCS} \}
\label{eq:sum}
\end{equation}
where $\phi_e$ is the scattered electron azimuthal angle and $\varphi$ is the 
out-of-plane angle between the leptonic and hadronic planes. The BH and DVCS 
processes are undistinguishable but the BH amplitude is completely known and 
exactly calculable from the electromagnetic form factors. Beam and/or target 
polarization degrees of freedom can be used to select different contributions 
to the cross section. Particularly, the polarized cross section difference 
for opposite beam helicities can be used to isolate the imaginary part of the 
DVCS amplitude, according to the expression~\cite{Gui98}
\begin{eqnarray}
\frac{d^5 \Delta \sigma}{dQ^2 dx_B dt d\phi_e d \varphi} & = & \frac{1}{2} 
\left[ \frac{d^5 \overrightarrow{\sigma}}{dQ^2 dx_B dt d\phi_e d \varphi} - 
\frac{d^5 \overleftarrow{\sigma}}{dQ^2 dx_B dt d\phi_e d \varphi} \right] 
\label{eq:dif} \\
& = & {\cal T}_{BH} \, \Im m \{{\cal T}_{DVCS}\} + \Re e \{{\cal T}_{DVCS}\} \,
\Im m \{{\cal T}_{DVCS}\} \nonumber
\end{eqnarray}
where $\Im m \{{\cal T}_{DVCS}\}$ appears now linearly, instead of 
quadratically, and magnified by the BH amplitude. 
 
\section{Nuclear Generalized Parton Distributions}

GPDs may be classified according to their helicity and chirality 
properties~\cite{{Die01},{Bof07}}. For a spin $S$ nucleus, one can define $(2S+1) 
\cdot (2S+1)$ parton helicity conserving and chiral even quark GPDs, and $(2S+1) 
\cdot (2S+1)$ parton helicity flipping and chiral odd quark GPDs. The same 
number of GPDs is needed for gluons, such that the complete partonic 
structure of the nucleus may be described by $4 \cdot (2S+1)^2$ parton GPDs, 
half of them conserving the nucleus helicity. DVCS is sensitive only to 
chiral even GPDs and is dominated by quark or gluon GPDs depending mainly on 
the studied $x_B$ region.

Similarly to nucleon GPDs, the optical theorem and the polynomiality 
constraints have remarkable consequences on nuclear 
GPDs~\cite{{Ber01},{Kir04}}. Considering for example a scalar nucleus target 
$A$, the quark and gluon GPDs are linked to the forward parton 
distributions following
\begin{eqnarray}
H^{f/A}(Q^2,x,\xi=0,t=0) & = & q^{f/A}(Q^2,x) \\
H^{f/A}_T(Q^2,x,\xi=0,t=0) & = & \delta q^{f/A}(Q^2,x) \\
H^{g/A}(Q^2,x,\xi=0,t=0) & = & x g^{A}(Q^2,x) 
\end{eqnarray}
where $q^{f/A}(Q^2,x)$, $\delta q^{f/A}(Q^2,x)$, and $g^{A}(Q^2,x)$ are the 
quark flavor $f$ density, transversity, and gluon distributions of the $A$ 
nucleus, respectively. The gluon heliciy-flip GPD is zero in the forward 
limit case. The first Mellin moment relates the quark GPDs to the nucleus 
electromagnetic form factor
\begin{equation}
\sum_f e_f \int^{1}_{-1} \, dx \, H^{f/A}(Q^2,x,\xi,t) = F^A(t) .
\end{equation}
A similar relation can be written for the quark chiral odd GPD
\begin{equation}
\sum_f e_f \int^{1}_{-1} \, dx \, H^{f/A}_T(Q^2,x,\xi,t) = \kappa^A_T(t)
\end{equation}
where $\kappa^A_T(t)$ represents the spin-flavour dipole moment~\cite{Bur05} of 
the isoscalar nucleus which may be accessed in electroproduction of 
neutral pions~\cite{Ahm08}. The second Mellin moment for quark and gluon 
GPDS writes 
\begin{equation}
\int^{1}_{-1} \, dx \, x H^{f/A}(Q^2,x,\xi,t) = M_2^{f/A}(t) + \frac{4}{5} \xi^2
d^{f/A}(t)
\end{equation}
where the first term of the right-hand side represents the momentum fraction
of the target carried by the quark, and the second term is the nuclear D-term
encoding information about the forces experienced by partons inside 
hadrons~\cite{Pol03}. In the particular case of a spin 1/2 nucleus, the D-term
dependence drops out in the sum of second Mellin moments of unpolarized 
chiral even GPDs. Its forward limit yields 
\begin{equation}
\frac{1}{2} \int^{1}_{-1} \, dx \, x { \left[ H^{f}(Q^2,x,\xi,0) + 
E^{f}(Q^2,x,\xi,0) \right] } = J^f = \frac{1}{2} \Delta \Sigma^f + L^f ,
\label{JiSR}
\end{equation}
the so-called Ji sum rule~\cite{Ji97} which allows access to the contribution
of the orbital momentum of partons to the spin of the nucleon. \newline
The previous relation (eq.~\ref{JiSR}) involves the $E^f$ GPD, a completely 
new distribution flipping the spin of the nucleus and therefore inaccessible 
to inclusive DIS experiments. Higher spin nuclei figure also several other 
unknown distributions related to peculiar features of the partonic structure 
of the nuclei~\cite{Ber01}.

\section{Coherent Deeply Virtual Compton Scattering}

\begin{figure}[t]\centering
\includegraphics[width=59mm]{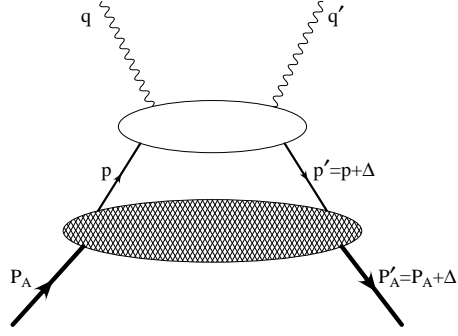}
\caption{Schematic diagram of the coherent DVCS reaction amplitude: the 
DVCS process occurs on a quark belonging to a nucleon embedded in the 
nuclear medium represented by a non-diagonal spectral function $S_F(p,p')
$~\cite{Liu05}.}
\label{fig:coh}
\end{figure}

Coherent DVCS corresponds to the channel of the DVCS reaction off nuclei 
where the final state consists exclusively of the initial nucleus 
(fig.\ref{fig:coh}). This process probes directly the nuclear GPDs i.e. the 
GPDs of the nucleus, that may be reconstructed from the elementary nucleon 
GPDs~\cite{{Kir04},{Guz03},{Liu05},{Guz08}} depending on specific models of 
the nuclear structure. As a consequence of the $t$-dependence of the nuclear
electromagnetic form factors, the coherent DVCS cross section is strongly
suppressed as the momentum transfer to the nucleus increases. This expected 
behaviour was experimentally observed on several nuclei at HERMES~\cite{Ell07} 
and on the deuterium at JLab~\cite{Maz07}. More dedicated exclusive 
experiments are required to obtain quantitative and conclusive information 
about nuclear GPDS. Particularly, the exclusivity constraint at small $t$ calls 
for the detection of the recoil nucleus. This is the strategy chosen by an
exploratory experiment on the $^4$He which aims to obtain the first
quantitative measurements of the imaginary and real parts of the $^4$He
Compton form factor~\cite {Egi08}.

\subsection{Elementary Case}

The $^4$He nucleus may be considered in many respects as an elementary 
system for the study of the impact of the nuclear medium on the partonic 
structure of nuclei: it is a well-known few-body system for which the most 
sophisticated microscopic calculations of the nuclear structure and dynamics 
are well-established~\cite{Car98}; it is a dense enough nucleus to generate 
sensitive effects on the parton distributions~\cite{Gee95}; in the context of 
the DVCS process, its leading twist partonic structure is described by one 
single Compton form factor ${\mathcal H}^A$ (sec.~3). 

The ratio of the beam polarized cross section difference (eq.~\ref{eq:dif}) 
and unpolarized cross section (eq.~\ref{eq:sum}) defines the beam spin 
asymmetry (BSA) measured for two opposite beam longitudinal helicities. For a 
spin zero nucleus, the BSA at the twist-2 accuracy writes
\begin{equation}
A_{LU}(\varphi) = \frac{\alpha_0(\varphi) \Im^A}{\alpha_1(\varphi) +
\alpha_2(\varphi) \Re^A + \alpha_3(\varphi)( {\Re^A}\cdot{\Re^A} + {\Im^A} 
\cdot{\Im^A} )} \label{eq:phi_fit}
\end{equation}
where $\Im^A = \Im m \{{\mathcal H}^A\}$, $\Re^A = \Re e \{{\mathcal H}^A\}$
are the unknown imaginary and real parts of the Compton form factor, and the 
$\alpha_i(\varphi)$'s are kinematical factors depending on $\varphi$. 
Therefore, measuring the $\varphi$ distribution of the BSA at a given 
kinematics $(Q^2,x_B,t)$ allows to extract simultaneously $\Im^A$ and 
$\Re^A$~\cite{Per08}. The simple BSA expression of eq.~\ref{eq:phi_fit} is a 
direct consequence of the spin zero of the $^4$He nucleus. Higher spin 
targets figure a combination of Compton form factors that does not factorize 
in a simple bilinear form~\cite{BelM02}. More strictly, gauge invariance of the DVCS
amplitude~\cite{Ani00} complicates this simple expression by introducing twist-3 
GPDs in the numerator and denominator of eq.~\ref{eq:phi_fit}~\cite{Bel01}. Twist-3
nuclear GPDs are mostly unknown distributions, and in the case of the $^4$He 
nucleus, one may expect to get some information from the BSA in the region of 
the minimum of the electromagnetic form factor, i.e. in a region where the BSA 
depends only on pure DVCS quantities.

\begin{figure}[t]\centering
\includegraphics[width=0.495\linewidth]{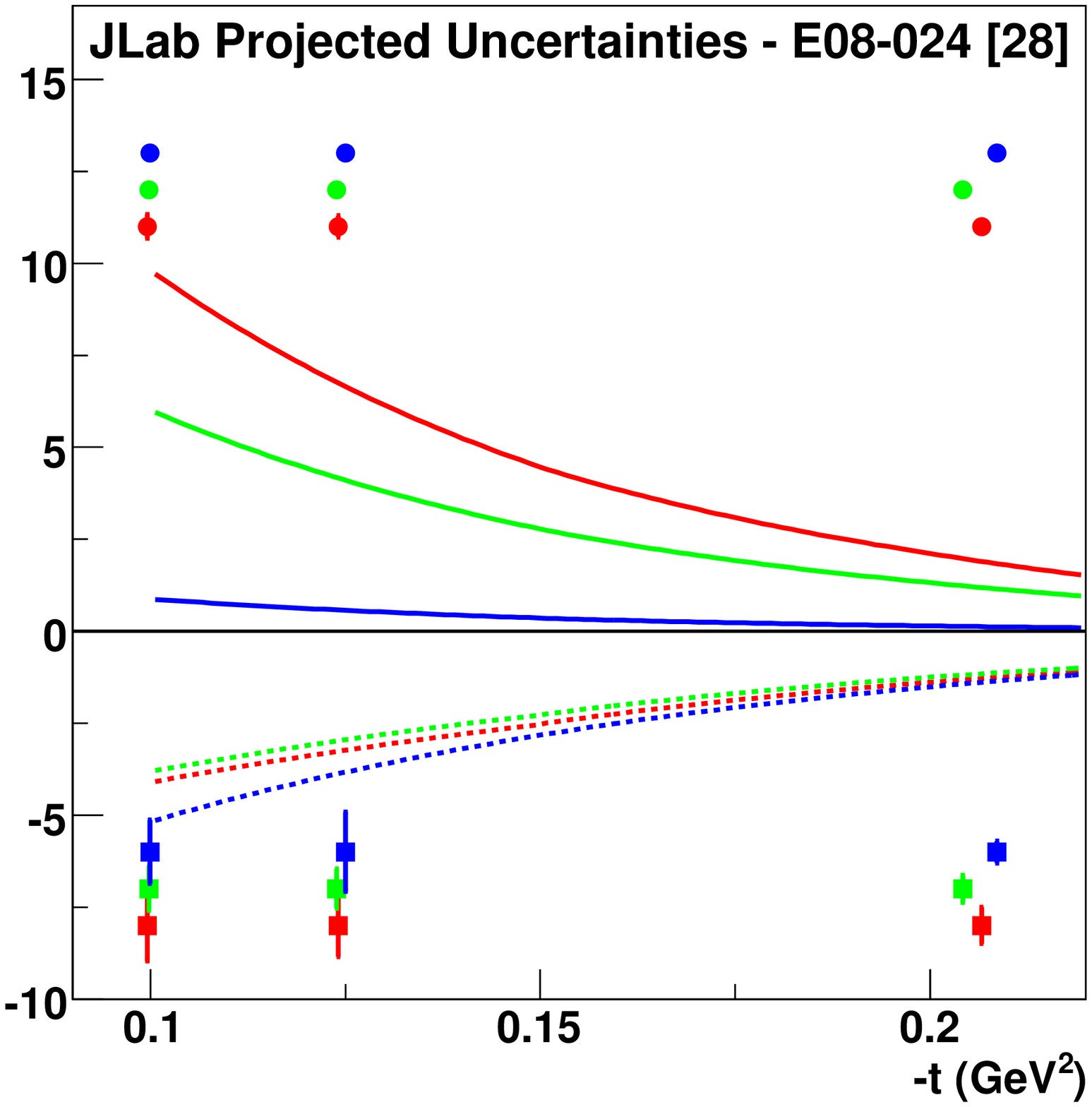}
\includegraphics[width=0.495\linewidth]{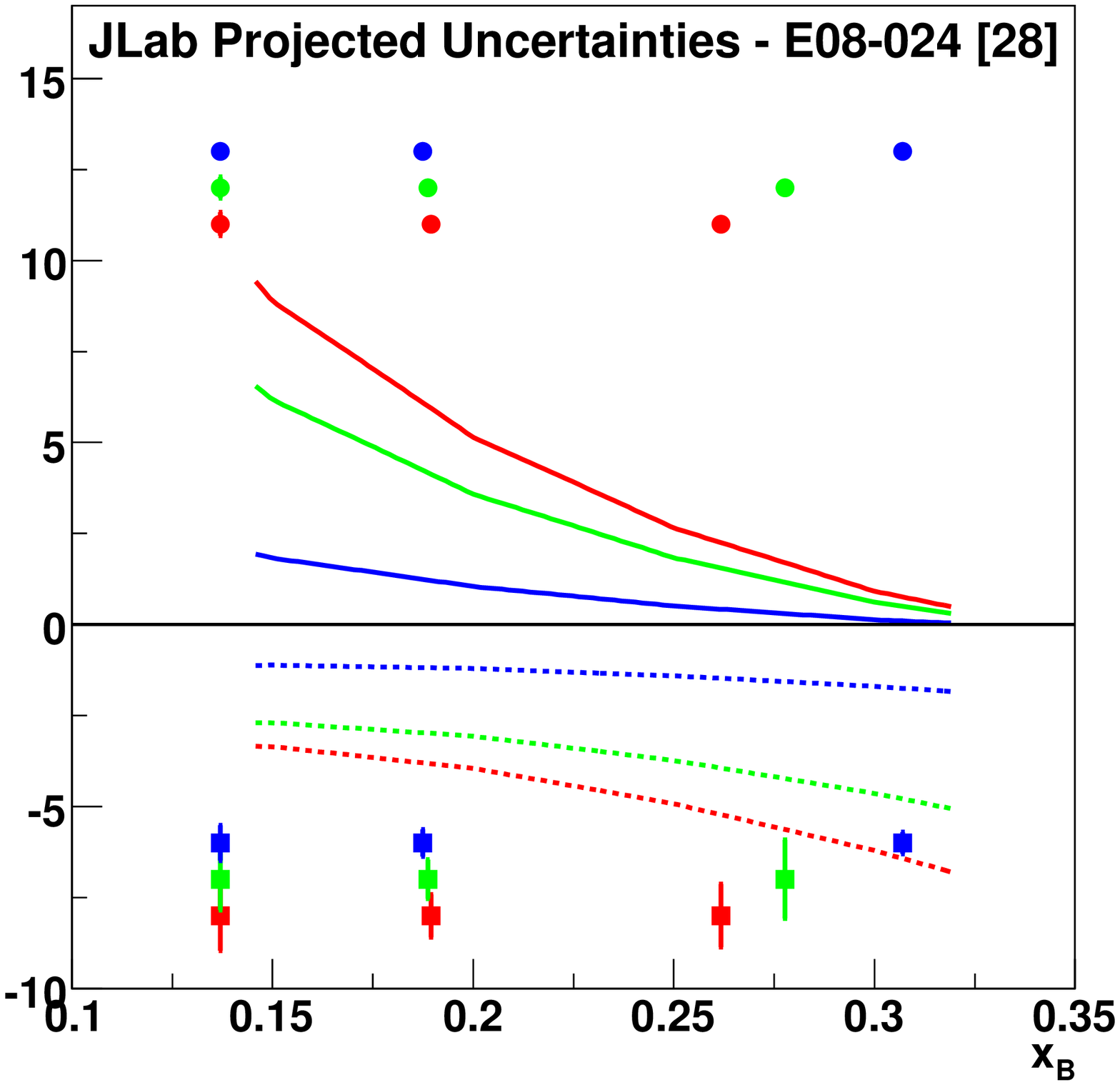}
\caption{Comparison of the projected uncertainties of the E08-024 DVCS 
experiment off $^4$He~\cite{Egi08} at different kinematics, to model 
predictions~\cite{Guz03}; the upper positive part of each panel represents
$\Im^A$ predictions, and the lower negative part, $\Re^A$'s; the blue, red, 
and green colors on the left panel correspond to $x_{B} = 0.13, \,0.19, \, 
0.28$, and on the right panel to $-t= 0.09, \, 0.12, \, 0.20$~GeV$^2$, 
respectively.}
\label{fig:proj_cff}
\end{figure}

Relying on this experimental technique, the E08-024 experiment at 
JLab~\cite{Egi08} will provide high quality data to control the importance of the
twist-3 contributions and extract the twist-2 $^4$He Compton 
form factor in the valence region (fig.~\ref{fig:proj_cff}). The exclusivity 
of the reaction channel will be ensured by the detection of the scattered
electrons with CLAS~\cite{Mec03}, the produced real photons with a specific
internal calorimeter at small angles ($4^{\circ} \le \theta_{\gamma}^{lab.} 
\le 15^{\circ}$) supplementing CLAS capabilities at larger angles, and the 
recoil nuclei with the BoNuS radial time-projection chamber~\cite{Fen08}. The 
$Q^2$, $x_B$, and $-t$ dependencies of ${\mathcal H}^A$ will be measured, 
however in a limited phase space. An exhaustive investigation of coherent
DVCS requires the full capabilities of the 12~GeV upgrade. The present 
experiment should provide a guide for these future studies of the
femto-tomography of the nucleus.
 
\subsection{Generalized EMC Ratio}

\begin{figure}[t]\centering
\includegraphics[width=0.57\linewidth]{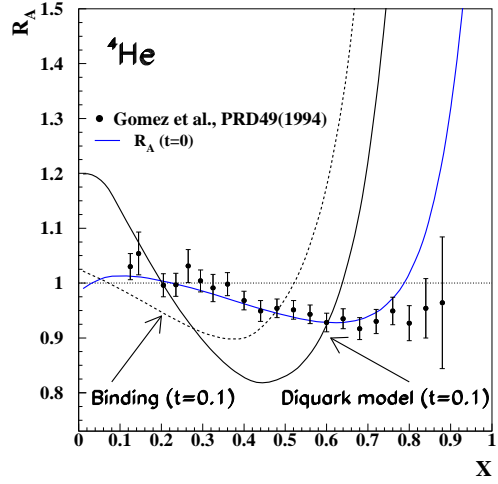}
\caption{Off-forward EMC effect in $^4$He at $-t=0.1$~GeV$^2$~\cite{Liu05:1}; 
theoretical calculations are shown for two different descriptions of the EMC 
effect, and are compared with experimental data~\cite{Gom94} and predictions 
for the forward case.}
\label{fig:ra}
\end{figure}

Nuclear effects are often expressed as the ratio of a given observable on a 
nuclear target with the same observable measured on an hydrogen target. The 
ratio $R_A$ of the amplitude of the nuclear and nucleon BSAs is then expected 
to be an indication of the nuclear medium effects on the partonic 
structure of the nucleus. In the case of an isoscalar target, and assuming 
the dominance of the BH cross section, this ratio may be written
\begin{eqnarray}
R_A (Q^2,x_B,t) & = & \frac{A_{LU}^{^4He}(Q^2,x_B,t,\varphi=\frac{\pi}{2})}
{A_{LU}^{^1H}(Q^2,x_B,t,\varphi=\frac{\pi}{2})} \\
& \propto & \frac{\Im m \left\{F^A {\mathcal H}^A \right\}}{\Im m \left\{ 
F_1{\mathcal H} + \xi (F_1+F_2) \widetilde{\mathcal H} - \frac{t}{4M^2} F_2
{\mathcal E} \right\}} \,
\frac{{\left|\tau_{BH}^{^1H}\right|}^2}{{\left|\tau_{BH}^{^4He}\right|}^2}
\\
& \approx & \frac{H^A}{H} \, \frac{F_1}{F^A} \label{eq:app}
\end{eqnarray}
where $F_{1(2)}$ is the Dirac(Pauli) nucleon electromagnetic form factor, and 
$\{ {\mathcal H}, \widetilde{\mathcal H}, {\mathcal E} \}$ are the nucleon Compton
form factors. The approximated expression of (eq.~\ref{eq:app})~\cite{Liu05:1} 
makes clear the physical meaning of $R_A$, especially in the forward limit 
where the GPDs reduce to the usual parton distributions: $R_A$ can be
interpreted as a generalized (or off-forward) EMC ratio, characterizing 
nuclear medium effects in the direction transverse to $\vec q$. The benefit 
of this ratio, supported by several 
calulations~\cite{{Guz08},{Liu05:1},{Sco04}}, is shown in fig.~\ref{fig:ra}. The 
generalized EMC ratio appears to be a very sensitive observable to the 
ingredients of the calculations i.e. the measurement of this observable should
help to disentangle unambiguously between the many scenarios proposed for the 
explanation of the EMC effect. In this calculation, the striking stronger
effects in the off-forward case suggest that the correlations between partons 
may play a crucial role in this problem.
   
\section{Incoherent Deeply Virtual Compton Scattering}

\begin{figure}[h]\centering
\includegraphics[width=51mm]{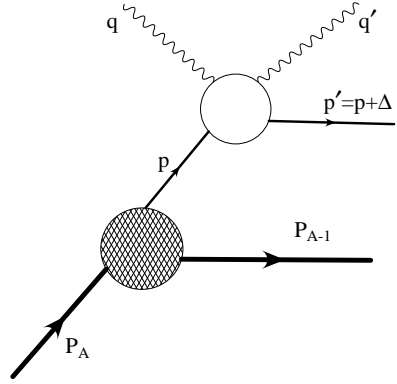}
\caption{Schematic representation of the incoherent DVCS process in the impulse
approximation: the elementary DVCS process is convoluted with the probability 
to find a nucleon with momentum $p$ and energy $E$ in the initial nucleus i.e.
the nuclear spectral function $S(p,E)$.}
\label{fig:inc}
\end{figure}

Incoherent DVCS is a complementary chanel to coherent DVCS where, in addition 
to the production of a real photon, a nucleon is expelled from the intial 
nucleus. In the impulse approximation, this process may be described as the 
interaction of the virtual photon with a quark belonging to a nucleon embedded 
in the nuclear medium (fig.~\ref{fig:inc}). The elementary GPD information is 
basically the same for coherent and incoherent DVCS, the essential difference 
arising from the nuclear information: incoherent DVCS involves the usual 
nuclear spectral function $S(p,E)$ while coherent DVCS considers a generalized 
skewed spectral function $S_F(p,p')$~\cite{{Liu05},{Sco04}}. Consequently,
incoherent DVCS can be considered as a direct way to access bound nucleon GPDs 
which quantify the effect of the nuclear medium on free nucleon 
GPDs~\cite{{Liu05:1},{Guz08:1}}.  

\subsection{Laboratory for Bound Nucleons}

The A$(e,e'p\gamma)$A-1 reaction isolates specifically partonic configurations 
of bound nucleons, and opens a new path for the investigation of bound nucleon 
properties. Here, the perpendicular component of the momentum transfer to the 
nucleon is the Fourier conjugate to the transverse position of partons inside 
the bound nucleon, an information required to access features like the 
confinement size of bound nucleons or the transverse overlap areas of hadronic 
configurations in the nucleus. Similarly to the coherent case, nuclear medium 
effects may be expressed in terms of the BSA ratio $R_{p/A}$ for the incoherent 
and free nucleon processes
\begin{eqnarray}
& & R_{p/A}(Q^2,x_B,t) = \frac{A_{LU}^{p/A}(Q^2,x_B,t,\varphi=\frac{\pi}{2})}
{A_{LU}^{^1H}(Q^2,x_B,t,\varphi=\frac{\pi}{2})} \\
& \propto & \frac{\Im m \left\{ F_1^{p/A}{\mathcal H^{p/A}} + \xi (F_1^{p/A}+
F_2^{p/A}) \widetilde{\mathcal H}^{p/A} - \frac{t}{4M^2} F_2^{p/A} {\mathcal E}^
{p/A} \right\}}{\Im m \left\{ F_1{\mathcal H} + \xi (F_1+F_2) \widetilde{\mathcal H} 
- \frac{t}{4M^2} F_2{\mathcal E} \right\}} \, \frac{{\left|\tau_{BH}^{^1H}\right|}
^2}{{\left|\tau_{BH}^{p/A}\right|}^2} \, \, \, \, \, \, \, \label{eq:rai}
\end{eqnarray}
where $F_{1(2)}^{p/A}$ is the Dirac(Pauli) bound nucleon electromagnetic form 
factor, and $\{ {\mathcal H}^{p/A}, \widetilde{\mathcal H}^{p/A}, {\mathcal E}^{p/A}
 \}$ are the bound nucleon Compton form factors. $R_{p/A}$ turns out to be a very 
sensitive observable to the details of the medium modification 
models~\cite{{Liu05},{Sco04}}. Future measurements of this observable off the 
$^4$He nucleus~\cite{Egi08} are strongly expected to help our understanding of 
bound nucleon properties.

However, as in the A$(e,e'p)$A-1 reaction, one should be attentive to the
effects of the initial Fermi motion and to the reaction mechanisms beyond impulse 
approximation. This is of particular importance for the BH process which is used 
as a reference light to reveal deviations originating from the DVCS mechanism. 
In this respect, the conservation of the electromagnetic current together with a 
reliable description of off-shellness effects are mandatory. Final state 
interactions between the recoil nucleon and the residual nucleus are also expected 
to impact the extraction of the DVCS signal from the distortion of the simple 
scheme of the impulse approximation via interferring amplitudes and imaginary BH. 
The magnitude of these several effects are currently investigated~\cite{Liu:Vou}.    

\subsection{Effective Neutron Target}

In abscence of free neutron targets, few-body systems are often used as effective 
neutron targets. Typically, the deuterium is considered as a free neutron target 
and polarized $^3$He as an effective polarized neutron target; higher mass nuclei
feature important contributions from other reaction mechanisms which make the
spectator nucleon picture unreliable~\cite{Mis02}.

\begin{figure}[t]\centering
\vspace*{7pt}
\includegraphics[width=69mm]{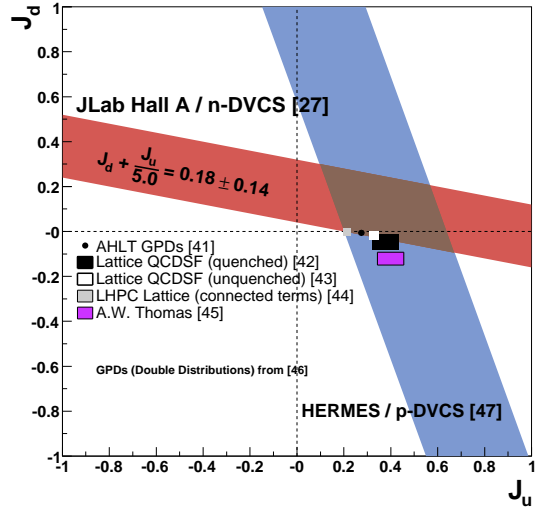}
\caption{Model dependent experimental constraint on $J_u$ and $J_d$ quark angular 
momenta from the JLab Hall A n-DVCS experiment~\cite{Maz07}. Different model 
calculations~\cite{{Ahm07},{Goc04},{Sch07},{Hag07},{Tho08}} are compared to an 
extrapolation of experimental data within the VGG double distribution description
of GPDs~\cite{Van99}. A similar constraint from the $\vec{\mathrm{p}}$-DVCS target
spin asymmetry measured by HERMES~\cite{Air08} is also shown.}
\label{fig:jujd}
\end{figure}

DVCS off the neutron appears as a process complementary to DVCS off the proton, 
and is a mandatory step in the hunt for the quark angular momentum~\cite{Maz07}. 
Indeed, the differences between neutron and proton electromagnetic form factors 
allow to access new combinations of GPDs. Furthermore, because of the electric 
charge weighting each flavor GPD (eq.~\ref{eq:cff}), the neutron is essentially 
sensitive to the $u$ quark flavor, similarly to the proton. Consequently to 
isospin symmetry, the measurement of neutron GPDs tells about the $d$ quark GPDs 
in the proton. This is exactly the message that was delivered by the JLab Hall A 
n-DVCS experiment~\cite{Ber03}. The complementary between neutron and proton data 
was expressed in terms of a model dependent extraction of the $u$ and $d$ quark
angular momenta (fig.~\ref{fig:jujd}) which shows that the beam polarized cross 
section difference off the neutron is, similarly to the transversely polarized 
target spin asymmetry, sensitive to the least known and constrained GPD 
$E$~\cite{Maz07} of particular importance in the angular momentum sum
rule~\cite{Ji97}. \newline
This last feature can be clearly understood from the inspection of the single
polarization DVCS observables for the neutron. In the small $t$ limit, the neutron 
Dirac form factor vanishes, as does the spin conserving polarized GPD $\widetilde{H}$ 
as a result of the cancelation between the $\Delta u$ and $\Delta d$ quark
helicity distributions. Within this approximation and at the twist-2 accuracy, the 
GPD dependent part~\cite{BelM02} of the cross section difference for a 
longitudinally polarized beam and an unpolarized target may be expressed
\begin{equation}
\frac{d^5 \overrightarrow{\sigma}}{d^5 \Omega} - 
\frac{d^5 \overleftarrow{\sigma}}{d^5 \Omega} \propto
- \frac{t}{4M^2} \, F_2 \, \Im m \left[ {\mathcal E} \right] \, \sin(\varphi)
\end{equation}
where $d^5 \Omega \equiv dQ^2 dx_B dt d\phi_e d \varphi$. Similarly, the GPD 
dependent part~\cite{BelM02} of the cross section difference for an unpolarized 
beam and a longitudinally polarized target may be written
\begin{equation}
\frac{d^5 {\sigma}^{\rightarrow}}{d^5\Omega} - 
\frac{d^5 {\sigma}^{\leftarrow}}{d^5\Omega} \propto
\xi \, F_2 \, \Im m \left[ {\mathcal H} + \frac{\xi}{1+\xi}{\mathcal E} - \frac{t}{4M^2} 
\widetilde{\mathcal E} \right] \, \sin(\varphi)  \, , \label{eq:ALT}
\end{equation}
and for a transversely polarized target with azimutal angle $\phi_S$
\begin{eqnarray}
\frac{d^5 {\sigma}^{\uparrow}}{d^5\Omega} - 
\frac{d^5 {\sigma}^{\downarrow}}{d^5\Omega} & \propto &
\left[ c_0^{DVCS} + c_0^I \right] \sin(\varphi-\phi_S) \\
& + & c_1^I \sin(\varphi-\phi_S) \cos(\varphi) + s_1^I \cos(\varphi-\phi_S) 
\sin(\varphi) \nonumber
\end{eqnarray}
where the coefficients of the harmonic development writes
\begin{eqnarray}
& & c_0^{DVCS} \propto (1+\xi) \, \Im m \left[ {\mathcal H}
{\mathcal E}^{\star} - {\mathcal E}{\mathcal H}^{\star} \right] \\
& & c_0^I \propto \frac{1}{1+\xi}  \frac{t}{2M^2} \frac{(2-y)^2}{1-y} F_2 \, \Im m \left[ 
\xi^2 \left( {\mathcal E} - \widetilde{\mathcal E} \right) -\left( 1-\xi^2 \right) {\mathcal H} 
\right] \label{eq:c0I} \\
& & \,\,\,\,\,\,\,\,\,\,\,\,\, + \frac{t}{M^2} F_2 \, \Im m \left[ {\mathcal H}
\right] \nonumber \\
& & c_1^I \propto \frac{1}{1+\xi}  \frac{t}{2M^2} F_2 \, \Im m \left[
\xi^2 \left( {\mathcal E} - \widetilde{\mathcal E} \right) - \left(1-\xi^2\right) {\mathcal H} 
\right] \label{eq:c1I} \\
& & s_1^I  \propto  \frac{2\xi}{1+\xi} F_2 \, \Im m \left[ \xi 
{\mathcal H} + \frac{\xi^2}{1+\xi} {\mathcal E} + \frac{t}{4M^2} \left( {\mathcal E} - \xi 
\widetilde{\mathcal E} \right) \right] . \label{eq:s1I}
\end{eqnarray}
In the JLab energy range $\xi$ remains smaller than 0.5, such that the dominant
contribution in eq.~\ref{eq:ALT}, eq.~\ref{eq:c0I}, eq.~\ref{eq:c1I}, and 
eq.~\ref{eq:s1I} is 
proportional to ${\mathcal H}$. For the transversely polarized target the 
sensitivity to $E$ appears through a pure DVCS harmonic coefficient, while the most
efficient single spin observable to access $E$ seems to be the polarized beam cross
section difference.  

\section{Conclusions}

DVCS off nuclei is a very promising tool for the investigation of the partonic
structure of nuclei. Preliminar measurements at HERMES and JLab have established 
the existence of a DVCS signal in beam spin asymmetries, but remain quantitatively 
limited. A full dedicated program starts to develop in the context of the JLab
upgrade at 12 GeV, and will benefit from the exploratory experiments on the 
$^2$H~\cite{Cam08} and $^4$He nuclei. Exclusivity is a key feature for the 
completion of this program. \newline
As today, the incoherent DVCS process is the main source of information for DVCS 
off the neutron. The perspective to use this channel to investigate the partonic 
structure of bound nucleons is motivating a lot of theoretical interest, but 
requires the next generation of experiments for quantitative and conclusive
information. The limits of the impulse approximation remains a matter of 
concern. 

\section*{Acknowledgments}

I would like to thank the organizers of the XXVIIth International Workshop on
Nuclear Theory for their invitation and warm hospitality at Rila Mountains. 
The development of a nuclear DVCS program at JLab is a collaborative work which 
benefits from many contributions. I would like to thank H.~Egiyan, F.-X.~Girod, 
V.~Guzey, K.~Hafidi, C.~Hyde, S.~Liuti, M.~Mazouz and B.~Pire for many stimulating 
and fruitfull discussions. 
   
This work was supported in part by the U.S. Department of Energy (DOE) 
contract DOE-AC05-06OR23177 under which the Jefferson Science Associates, 
LLC, operates the Thomas Jefferson National Accelerator Facility, the 
National Science Foundation, the French Atomic Energy Commission and National 
Center of Scientific Research, and the GDR n$^{\circ}$3034 Physique du
Nucl\'eon.


\begin{thebibliography}{99}
 
\bibitem{Ji97}
X.~Ji, {\em Phys. Rev. Lett.} {\bf 78} (1997) 610.
 
\bibitem{Air01}
A.~Airapetian {\it et al.}, {\em Phys. Rev. Lett.} {\bf 87} (2001) 182001.
 
\bibitem{Ste01}
S.~Stepanyan {\it et al.}, {\em Phys. Rev. Lett.} {\bf 87} (2001) 182002.

\bibitem{Mun06}
C.~Mu\~noz Camacho, A.~Camsonne, M.~Mazouz, C.~Ferdi, G.~Gavalian, E.~Kuchina
{\it et al.}, {\em Phys. Rev. Lett.} {\bf 97} (2006) 262002.

\bibitem{Gir08}
F.-X.~Girod, R.A.~Niyazov {\it et al.}, {\em Phys. Rev. Lett.} {\bf 100} (2008) 
162002.

\bibitem{Mul94}
D.~M\"uller, D.~Robaschick, B.~Geyer, F.M.~Dittes, J.~Ho{\v r}ej{\v s}i, {\em
Fortschr. Phys.} {\bf 42} (1994) 101.
 
\bibitem{Rad97}
A.V.~Radyushkin, {\em Phys. Rev.} {\bf D 56} (1997) 5524.
 
\bibitem{Bur00}
M.~Burkardt, {\em Phys. Rev.} {\bf D 62} (2000) 071503(R).
 
\bibitem{Ral02}
J.P.~Ralston, B.~Pire, {\em Phys. Rev.} {\bf D 66} (2002) 111501(R).
 
\bibitem{Die02}
M.~Diehl, {\em Eur. Phys. Jour.} {\bf C 25} (2002) 223. 

\bibitem{Bel02}
A.V.~Belitsky, D.~M\"uller, {\em Nucl. Phys.} {\bf A 711} (2002) 118c.
 
\bibitem{Die03}
M.~Diehl, {\em Phys. Rep.} {\bf 388} (2003) 41.

\bibitem{Aub83}
J.J.~Aubert {\it et al.}, {\em Phys. Lett.} {\bf B 123} (1983) 275.

\bibitem{Ji98}
X.~Ji, J.~Osborne, {\em Phys. Rev.} {\bf D 58} (1998) 094018.
 
\bibitem{Col99}
J.C.~Collins, A.~Freund, {\em Phys. Rev.} {\bf D 59} (1999) 074009.
 
\bibitem{Gui98}
P.A.M.~Guichon, M.~Vanderhaeghen, {\em Prog. Part. Nucl. Phys.} {\bf 41} (1998)
125.
 
\bibitem{Die01}
M.~Diehl, {\em Eur. Phys. J.} {\bf C 19} (2001) 485.

\bibitem{Bof07}
S.~Boffi, B.~Pasquini, {\em Riv. Nuo. Cim.} {\bf 30} (2007) 387.

\bibitem{Ber01}
E.R.~Berger, F.~Cano, M.~Diehl, B.~Pire {\em Phys. Rev. Lett.} {\bf 87} (2001) 
142302.

\bibitem{Kir04}
A.~Kirchner, D.~M\"uller, {\em Eur. Phys. J.} {\bf C 32} (2004) 347.

\bibitem{Bur05}
M.~Burkardt, {\em Phys. Rev.} {\bf D 72} (2005) 094020.

\bibitem{Ahm08}
S.~Ahmad, G.R.~Goldstein, S.~Liuti, {\em ArXiv:hep-ph} {\bf 0805.3568} (2008).

\bibitem{Pol03}
M.~Polyakov, {\em Phys. Lett.} {\bf B 555} (2003) 57.

\bibitem{Guz03}
V.~Guzey, M.~Strikman, {\em Phys. Rev.} {\bf C 68} (2003) 015204.

\bibitem{Liu05}
S.~Liuti, S.K.~Taneja, {\em Phys. Rev.} {\bf C 72} (2005) 032201(R).
 
\bibitem{Guz08}
V.~Guzey, {\em ArXiv:hep-ph} {\bf 0801.3235} (2008). 

\bibitem{Ell07}
F.~Ellinghaus, {\em ArXiv:hep-ex} {\bf 0710.5768} (2007).

\bibitem{Maz07}
M.~Mazouz, A.~Camsonne, C.~Mu\~noz Camacho, C.~Ferdi, G.~Gavalian, 
E.~Kuchina {\it et al.}, {\em Phys. Rev. Lett.} {\bf 99} (2007) 242501.
 
\bibitem{Egi08}
H.~Egiyan, F.-X.~Girod, K.~Hafidi, S.~Liuti, E.~Voutier {\it et al.}, {\em
JLab Proposal} {\bf E08-024} (2008).

\bibitem{Car98}
J.~Carlson, R.~Schiavilla, {\em Rev. Mod. Phys.} {\bf 70} (1998) 743.

\bibitem{Gee95}
D.F.~Geesaman, K.~Saito, A.W.~Thomas, {\em Ann. Rev. Nucl. Part. Sci.} 
{\bf 45} (1995) 337.

\bibitem{Per08}
Y.~Perrin, Master Thesis, Universit\'e Joseph Fourier, Grenoble (France),
2008.

\bibitem{BelM02}
A.V.~Belitsky, D.~M\"uller, A.~Kirchner, {\em Nucl. Phys.} {\bf B 629} (2002)
323.

\bibitem{Ani00}
I.V.~Anikin, B.~Pire, O.V.~Terayev, {\em Phys. Rev.} {\bf D 62} (2000) 071501.

\bibitem{Bel01}
A.V.~Belitsky, D.~M\"uller, A.~Kirchner, A.~Sch\"afer, {\em Phys. Rev.} 
{\bf D 64} (2001) 116002.

\bibitem{Mec03}
B.A.~Mecking {\it et al.}, {\em Nucl. Inst. Meth.} {\bf A 503} (2003) 513.

\bibitem{Fen08}
H.~Fenker {\it et al.}, {\it Nucl. Inst. Meth.} {\bf A 592} (2008) 273.
 
\bibitem{Liu05:1}
S.~Liuti, S.K.~Taneja, {\em Phys. Rev.} {\bf C 72} (2005) 034902.
 
\bibitem{Sco04}
S.~Scopetta, {\it Phys. Rev.} {\bf C 70} (2004) 015205.

\bibitem{Gom94}
J.~Gomez {\it et al.}, {\em Phys. Rev.} {\bf D 49} (1994) 4348.

\bibitem{Guz08:1}
V.~Guzey, A.W.~Thomas, K.~Tsushima, {\em ArXiv:hep-ph} {\bf 0806.3288} (2008).

\bibitem{Liu:Vou}
S.~Liuti, E.~Voutier, {\it work in progress}.

\bibitem{Mis02}
A.~Misiejuk, Z.~Papandreou, E.~Voutier {\it et al.}, {\em Phys. Rev. Lett.} {\bf 89} 
(2002) 172501.

\bibitem{Ahm07}
S.~Ahmad, H.~Honkanen, S.~Liuti, S.K.~Taneja, {\em Phys. Rev.} {\bf D 75} (2007)
094003.

\bibitem{Goc04}
M.~G\"ockeler {\it et al.}, {\em Phys. Rev. Lett.} {\bf 92} (2004) 042002.

\bibitem{Sch07}
G.~Schierholz, Proc. of the Workshop on Exclusive Reactions at High Momentum
Transfer, Jefferson Laboratory, Newport News (VA, USA), May 21-24, 2007.

\bibitem{Hag07}
Ph.~H\"agler {\it et al.}, {\em ArXiv:hep-lat} {\bf 0705.4295} (2008).

\bibitem{Tho08}
A.W.~Thomas, {\em ArXiv:hep-ph} {\bf 0803.2775} (2008).

\bibitem{Van99}
M.~Vanderhaeghen, P.A.M.~Guichon, M.~Guidal, {\em Phys. Rev.} {\bf D 60} (1999)
094017.

\bibitem{Air08}
A.~Airapetian {\it et al.}, {\em ArXiv:hep-ex} {\bf 0802.2499} (2008).

\bibitem{Ber03}
P.Y.~Bertin, C.~Hyde, F.~Sabati\'e, E.~Voutier {\it et al.}, {\em JLab Proposal} 
{\bf E03-106} (2003).

\bibitem{Cam08}
A.~Camsonne, C.~Hyde, M.~Mazouz {\it et al.}, {\em JLab Proposal} {\bf E08-025} (2008).

\end{thebibliography}
\end{document}